\documentclass[10pt]{article}
\usepackage{amssymb}
\usepackage{amsmath}
\usepackage{amstext}
\def \RR {{\mathbb R}}      
\def \ZZ {{\mathbb Z}}      
\def \NN {{\mathbb N}}      
\def \inv {^{-1}}
\def \eps {\varepsilon}
\def\wick#1{\hbox{:}#1 \hbox{:}}
\def\ket#1{\vert#1\rangle}
\def\RA {\Rightarrow}  

\renewcommand{\geq}{\geqslant}
\renewcommand{\leq}{\leqslant}
\def \bea {\begin{eqnarray}} \def \eea {\end{eqnarray}} 
\def\be{\begin{equation}} \def\ee{\end{equation}}

\begin{document}

\title{Conformal quantum field theory in various dimensions\thanks{to
    appear in: Proceedings ``Algebraic Methods in Quantum Field Theory'', Sofia, 2009,
    Heron Press (Sofia)}}  
\author{Marcel Bischoff, Daniel Meise, Karl-Henning Rehren, Ingo
  Wagner \\[1cm] {\small Institut f\"ur Theoretische Physik, Universit\"at G\"ottingen,} \\ {\small Friedrich-Hund-Platz 1, D-37077 G\"ottingen, Germany}
}

\maketitle

\begin{center}\it Dedicated to Ivan Todorov on the occasion of his
  75th birthday 
\end{center}

\parskip0.5mm

\begin{abstract}
Various relations between conformal quantum field theories in one, two
and four dimensions are explored. The intention is to obtain a better
understanding of 4D CFT with the help of methods from lower
dimensional CFT.
\end{abstract}

PACS 2008: {03.70.+k,11.10.-z}

\sloppy \raggedbottom

\setcounter{page}{1}

\section{Introduction}\label{sec1}
Quantum field theory (QFT) in four spacetime dimensions (4D) continues to be a
great challenge after many decades of intense research. While
perturbative and nonperturbative approximation schemes have proven
most efficient for many purposes, the rigorous construction of
nontrivial theories has not been achieved.  

On the other hand, in two dimensions (2D), many nontrivial models have been
constructed. A huge body of model-independent knowledge has been accumulated in
particular for conformal field theories; depending on the value of the
central charge, there are even classifications available. 

To close the gap between 2 and 4 dimensions, one would like to be able
to transfer general knowledge from 2 to 4. This is (besides its
manifold statistical mechanic applications) the main raison d'\^etre
for the study of lowdimensional models. In this contribution, we
present a number of attempts hopefully leading to new insights into 
the structure of correlation functions in nontrivial 4D conformal
QFT which is admissible from an axiomatic point of view.

To have the maximum power of this approach available, we assume the
strongest form of conformal symmetry, called ``global conformal
invariance'' (GCI) in \cite{NT01,NRT05}: 
the conformal group is implemented by a {\em true} representation on the
Hilbert space. This implies that the covariant fields have integer
scaling dimensions and satisfy Huygens' principle, i.e., they commute
not only at spacelike but also at timelike distance. Moreover, their
correlation functions are rational, and in fact polynomial after
multiplication with sufficiently high powers of Lorentz square distances
$\rho_{ij}=(x_i-x_j)^2$. While these features are conspicuously close
to free field theory, we shall indicate below why we expect nontrivial
fields within this highly restricted class. Notice that the massless
free field in $D>2$ dimensions has scaling dimension $\frac{D-2}2$, so
this field does not satisfy GCI if $D$ is odd. Recall also that in $D=2$, 
the massless free field does not exist because it is too singular at
zero momentum, but its gradient $j_\mu=\partial_\mu\varphi$ can be
defined. It is a conserved vector current of scaling dimension 1 and
decomposes into two chiral fields $j_0(x)\pm j_1(x)=j_\pm(x^0\pm x^1)$.   

In $D\geq 4$ even spacetime dimensions, GCI proved to be a highly
restrictive symmetry. In Sect.\ \ref{sec2}, we discuss the leeway
it allows beyond free fields in terms of the pole structure of correlation
functions. Remarkably, the new features can arise only in at least
six-point correlations -- which are hardly ever studied! 

The main open question is, whether this leeway is compatible with
Hilbert space positivity. A powerful method to approach this question
for four-point correlations is the partial wave expansion;
unfortunately, the partial waves are not known for more than four
points. Part of the subsequent sections
about ``restriction'' is motivated by attempts to find alternative
approaches to positivity applicable to higher correlation functions,
to which we return in Sect.\ \ref{sec5}.

\section{Conservation laws}\label{sec2}
\setcounter{equation}{0}

\subsection{Conserved tensor fields} 

Consider conformal symmetric traceless tensor fields of rank $r$ and
scaling dimension $d$. The quantity $d-r$ is called ``twist''. The
fields of twist $D-2$, where $D$ is the spacetime dimension, are
distinguished: their two-point function is determined by conformal
invariance and turns out to have zero divergence. By the
Reeh-Schlieder theorem it follows that these fields are conserved
tensor fields:  
\be\label{cl}\partial_\mu T^{\mu\ldots\nu}=0,\ee
with the exception of $r=0$, $d=D-2$. Except for the scalars, the
twist $D-2$ fields have the lowest possible dimension for the given
tensor rank admitted by the unitarity bound \cite{M77}. The scalars
have been proven, in $D=4$ dimensions \cite{NRT08}, to be either Wick
squares of massless free fields, or generalized free fields. We expect
that a similar argument holds also in $D=2n=6,8,\dots$ dimensions for
scalar fields of scaling dimension $D-2$. (This cannot be true for odd
$D$ because the massless free field violates GCI, and also not for
$D=2$ because the massless free field does not exist.)  

In $D=2$, the distinguished fields are precisely those which decompose
into chiral fields: Symmetric traceless tensors have only two
independent components $T_{++\ldots+}$ and $T_{--\ldots-}$, and by the
conservation law, these depend only on $x^0\pm x^1$. Indeed, almost
all our knowledge about 2D CFT relies on the presence of these
distinguished chiral fields, such as currents or the stress-energy tensor.

\subsection{Biharmonic fields} \label{bih}
Also in even dimension $D\geq 4$, the presence of conserved tensor
fields has far reaching consequences. For definiteness, $D=4$
throughout this section, although the statements generalize to even
$D\geq4$. The operator product expansion of any pair of fields $A$ 
and $B$ can be organized according to the twist of the composite
fields. If $A$ and $A'$ are scalar of equal dimension $d$, then the
lowest contribution to $A'(x)A(y)$ after the vacuum contribution is
that of twist $D-2$. Multiplying this contribution by  
$((x-y)^2)^{d-1}$, one arrives at a ``bifield'' $V_{A'A}(x,y)$, while
all higher twist contributions are of higher order in $(x-y)^2$. 
The infinitely many conservation laws for the local fields comprised
in $V_{A'A}$ can be cast into the simple form, called ``biharmonicity''
\cite{NST02}: 
\be\label{bih}\square_x V_{A'A}(x,y) = 0 = \square_y V_{A'A}(x,y).\ee

Biharmonicity is a highly nontrivial feature. By a classical
result \cite{BT}, every power series $p$ in $z\in\RR^n$ has a unique ``harmonic
completion'' $h=p+z^2\cdot q$, such that $h$ is harmonic: $\square_z
h=0$, and $q$ is another power series. But correlation functions
involving $V_{A'A}(x,y)$ are harmonic both w.r.t.\ $x$ and w.r.t.\ $y$. 
Therefore, the contribution from $V_{A'A}(x,y)$ in a correlation functions
involving $A'(x)A(y)$ must coincide with two a priori different harmonic
completions (w.r.t.\ $z=x-y$). The condition that the two completions
coincide is found, for purely scalar correlations, to be a universal
third order linear partial differential equation to be satisfied by
the function $U_0$ defined by  
$$\big\langle \cdots \big[A'(x)A(y)-\langle A'(x)A(y)\rangle\big] \cdots\big\rangle = 
\frac 1{((x-y)^2)^{d-1}}\big(U_0 + O((x-y)^2)\big),$$
where $U_0$ is a Laurent polynomial in the Lorentz square distances $\rho_{xi} =
(x-x_i)^2$, $\rho_{yi} = (y-x_i)^2$, and $\rho_{ij} =\rho_{ji} = (x_i-x_j)^2$
($x_i$ are the coordinates of the other scalar fields in the
correlation), homogeneous of degree $-1$ in both sets of variables
$\rho_{xi}$ and $\rho_{yi}$. When this condition is fulfilled, the
contribution from $V_{A'A}$ to the above correlation is the unique
biharmonic completion $V_0$ of $U_0$:
$$\langle \cdots V_{A'A}(x,y) \cdots\rangle = U_0 + O((x-y)^2).$$

The PDE to be satisfied by $U_0$ reads 
\be\label{pde}
\Big[\big(\sum_i\rho_{yi}\partial_{\rho_{xi}}\big)\big(\sum_{i <
  j}\rho_{ij}\partial_{\rho_{yi}}\partial_{\rho_{yj}}\big)-(x\leftrightarrow
y)\Big] U_0=0.\ee
Together with rationality, it is highly restrictive and
constrains the admissible form of $U_0$ far beyond conformal
invariance. In \cite{NRT08}, it was shown that the only poles of $U_0$
in the arguments $x,y\in\RR^4$ can be of the form
$$\frac P {\rho_{xm}^a\rho_{xn}^b\cdot
  \rho_{yn}^c\rho_{ym}^d}\qquad\hbox{with}\quad a,b,c,d\geq 0, $$
for some pair $m\neq n$, where $P$ is polynomial in $\rho_{xi}$ and
$\rho_{yi}$ ($i\neq m,n$), and a Laurent polynomial in all other
squared Lorentz distances $\rho_{ij}$. We call this structure a
``double pole'' if both $a$ and $b$ are positive, or if both $c,d >0$.   

The relevance of this observation is the following: Free field
examples of biharmonic fields are $\wick{\varphi(x)\varphi(y)}$ and
$\wick{\overline\psi(x)(x_\mu-y_\mu)\gamma^\mu\psi(y)}$, where
$\varphi$ and $\psi$ are the free massless scalar and Dirac
field. But correlation functions of Wick products of free fields and
their derivatives can only produce ``single poles'' with $a=0$ or
$b=0$, and $c=0$ or $d=0$.  Therefore, any double pole is a clear
signal of a nontrivial GCI CFT. On the other hand, double poles cannot
arise in four-point functions just because there are not sufficiently
many variables. Therefore, this signal can only be seen in at least
five-point correlations \cite{NRT08}.   

An example of a six-point double pole structure was presented in
\cite{NRT08}. A more systematic study was made by one of us \cite{MB}.
For a double pole as above, we call $a+b+c+d$ its order. A double pole
structure (DPS) is a rational solution to the PDE (\ref{pde}) 
$$\sum_{a,b,c,d}\frac {P_{abcd}} {\rho_{xm}^a\rho_{xn}^b\cdot
  \rho_{yn}^c\rho_{ym}^d}$$
involving nonzero terms with $a$ and $b>0$, or $c$ and $d>0$. Their polynomial
(in $\rho_{xi}$ and $\rho_{yi}$, $i\neq m,n$) coefficient functions
turn out to be organized into multiplets of $sl(2)$, whose generators are the
differential operators   
$$2H=\sum_{i\neq m,n}\rho_{xi}\partial_{\rho_{xi}} -
  \rho_{yi}\partial_{\rho_{yi}}, \quad X= \sum_{i\neq m,n}
    \rho_{xi}\partial_{\rho_{yi}},\quad Y=\sum_{i\neq
      m,n}\rho_{yi}\partial_{\rho_{xi}}.$$
More precisely, every DPS is a linear combination of DPSs obtained as
follows. Fix a pair of indices $m,n$. Fix four integers $0\leq p<a$,
$0\leq q<b$. Let $\ell=p+q$ and choose a monomial $P_\ell$ of order
$\ell$ in the variables $\rho_{xi}$ ($i\neq m,n$). $P_\ell$ is then a
highest weight vector of $sl(2)$: $H P_\ell= \frac\ell 2 P_\ell$ and  
$X P_\ell=0$. Let $k=a+b-\ell-1\geq 1$ and choose a monomial
$Q_k$ of order $k$ in the $sl(2)$ singlet variables
$R_{ij}=\rho_{xi}\rho_{yj}-\rho_{xj}\rho_{yi}$ ($i,j\neq m,n$). Notice 
that for five-point correlations, such singlets are not available,
hence one can also exclude five-point DPSs. These data, together with
a Laurent monomial $L$ in the variables $\rho_{kl}$ so as to saturate the
scaling dimension of the scalar fields in the correlation function,
induce DPSs of maximal order $\mu = 2(a+b)-\ell$, whose double poles
of order $=\mu$ are given in closed form by   
$$\sum_{\delta=0}^p\sum_{\eps=0}^q\frac{\rho_{xm}^\delta\rho_{xn}^\eps\cdot\rho_{ym}^{q-\delta}\rho_{yn}^{p-\eps}}{\rho_{xm}^a\rho_{xn}^b\cdot\rho_{ym}^b\rho_{yn}^a}\cdot 
\frac{(b-q)_\delta(-p)_\delta}{(1-a)_\delta\,\delta!}\frac{(a-p)_\eps(-q)_\eps}{(1-b)_\eps\,\eps!}\cdot\Big\vert\frac\ell2,\frac\ell2-\delta-\eps\Big\rangle\cdot
Q_k\cdot L$$
where the functions $\ket{\frac\ell2,\frac\ell2-\nu} =
(-1)^\nu(-\ell)_\nu Y^\nu P_\ell$ are vectors of weight $\frac\ell
2-\nu$  in the irreducible highest weight $sl(2)$ module generated by
$P_\ell=\ket{\frac\ell2,\frac\ell2}$. These contributions exhaust a
twodimensional rectangular sublattice within the lattice
$a+b+c+d=\mu$. The poles of order $<\mu$ are then determined
recursively from those of maximal order $=\mu$, because equation
(\ref{pde}) connects different orders. The system is in fact   
overdetermined, but in all cases studied it could be solved. We
conjecture that this is always the case. The solution is unique up to
DPSs of lower maximal order.  

Once the solution $U_0$ to (\ref{pde}) is given, its biharmonic
completion, i.e., the corresponding correlation function $\langle
\cdots V_{A'A}(x,y) \cdots\rangle$ solving (\ref{bih}), can be computed
recursively as a power series in $(x-y)^2$. Unlike the correlations
of local fields, these correlations are always transcendental
functions if $U_0$ contains double poles. In this case, $V_{A'A}$
cannot be Huygens bilocal, but is presumably Einstein bilocal in
general, as a case study in \cite{Varna} indicates.

\section{Restrictions}\label{sec3} 
\setcounter{equation}{0}

\subsection{Timelike surfaces}
The restriction of a quantum field to a timelike hypersurface yields
another Wightman field in lower dimensions \cite{Bo}. In this way, 
4D fields give rise to 3D and to 2D fields. It is also known that
conformal fields restrict to conformal fields on the hypersurface, and
the decomposition of conformal tensor fields can be described in terms
of ``internal derivatives'' of the original fields \cite{BT,DMPPT,M}. 

One can ultimately restrict the field to the time axis: because of
Huygens locality, this yields a local conformal 1D field depending
only on $x^0$. Notice that this step is quite different from the
decomposition of conserved 2D tensor fields into their chiral
components, that depend only on $x^0\pm x^1$. Yet, in both cases one
arrives at M\"obius covariant chiral fields!  

To give an example: The correlation functions of restricted fields are
just the restrictions of the original correlation functions. In
particular, free fields remain free in the sense that the truncated
correlations remain zero. Thus, if we restrict the massless free field
$\varphi$ in $D=4$ to the plane $x^2=x^3=0$, we arrive at a
generalized free field with the two-point function 
$$D(x-y)\vert_{\RR^2}=\frac{(2\pi)^{-2}}{(x^1-y^1)^2-(x^0-y^0-i\eps)^2}$$ 
of scaling dimension $d=1$. But because the spacetime dimension has
changed, its K\"allen-Lehmann weight is no longer a $\delta$-function at
$m^2=0$ but a continuum of all masses integrated with the measure
$dm^2$. Such fields do not possess a stress-energy tensor as a
Wightman field, because its two-point function diverges
\cite{DR}. Formally, one may assign an ``infinite central charge'' to
this SET. One may actually represent the generalized free field in 2
dimensions as a ``central limit'' $n\to\infty$ of     
$$\varphi_n(x)= n^{-\frac12}\sum_{\nu=1}^n\psi_\nu(x^0+x^1) \otimes \psi_\nu(x^0-x^1)$$
where $\psi_\nu$ are $n$ independent chiral real free Fermi fields,
hence the SET for $\phi_n$ has central charge $c= \frac n2\to\infty$. 

On the other hand, restricting $\varphi$ to the time axis, its two-point
function is just 
$$D(x-y)\vert_{\RR} = (2\pi)^{-2}\Big(\frac {-i}{x^0-y^0-i\eps}\Big)^2,$$
the two-point function of a canonical chiral current $j(x^0)$. The Wick
square $\wick{\varphi^2(x)}$ restricts to $\wick{j^2(x^0)} = 
\pi\inv T(x^0)$, where $T$ is the chiral stress-energy tensor with $c=1$.

\subsection{Null surfaces} \label{null}

A different option is the restriction to null hypersurfaces such as
$N=\{x\in\RR^4: x^0=x^1\}$. This case is not covered by the result in
\cite{Bo}. Yet, {\em massive} free scalar fields can be restricted. 
More precisely, the naive restriction has an infrared singularity,  
which can be cured by taking a derivative w.r.t. $x_+$,
where $x_\pm= x^0\pm x^1$. Then, defining 
$\varphi_N(x_+,x_\perp):=\partial_+\varphi_m(x)\vert_{x_-=0}$, one computes
\be\label{lf2}
\langle \varphi_N(x_+,x_\perp)\varphi_N(y_+,y_\perp)\rangle = \frac 1{4\pi}\,
\delta(x_\perp-y_\perp) \cdot\Big(\frac {-i}{x_+-y_+-i\eps}\Big)^2.\ee
This restriction is an instance of the more general situation studied
in \cite{KW}. The result is nothing but an infinite 
system of canonical free currents 
$j_n(x_+)=\int d^2 x_\perp \,\varphi_N(x_+,x_\perp)\,f_n(x_\perp)$,
where $f_n$ is an orthonormal basis of $L^2(\RR^2)$. The remarkable
fact is that the vacuum fluctuations associated with the transverse
coordinates $x_\perp\in\RR^2$ are completely suppressed \cite{S1}, and
these degrees of freedom are traded into an infinite-dimensional inner
symmetry. Moreover, the restriction is {\em independent of the
  original mass}. 

Looking at the field as a distribution, the construction means that
the extension to test functions of the form $f(x_+,x_\perp)\delta(x_-)$ 
must be bought by the constraint that $f=\partial_+g$ where $g$ is a
test function on $\RR^3$. Because the restriction is independent of
the mass, {\em every} scalar two-point function restricts to the same
result (\ref{lf2}) times the integral over the K\"allen-Lehmann
density. In particular, two-point functions of scalar fields where
this integral is divergent cannot be restricted in the same way, such
as the Wick square or non-superrenormalizable interacting
fields. Moreover, the derivatives $\partial_+$ do not properly cure
at the same time the single contraction terms appearing in higher
correlation functions.   

However, one can restrict the bifield 
$\wick{\varphi_m(x)\varphi_m(y)}$ via 
$$\partial_{x_+}\partial_{y_+}\wick{\varphi(x)\varphi(y)}\vert_{x_-=y_-=0}
= \wick{\varphi_N(x_+,x_\perp)\varphi_N(y_+,y_\perp)}.$$
(One may then well pass to coinciding points $x_+=y_+$ {\em after} taking the
derivatives and smearing in the transversal space $\RR^2$, but this is
obviously not an operation on the Wick square itself.)  

For $m=0$, the bifield $\wick{\varphi(x)\varphi(y)}$ is the simplest 
instance of a biharmonic field, as discussed in the previous section. 
This suggests a speculation that biharmonic fields can always be
restricted. This expectation is supported by the solution to the
characteristic initial value problem for the wave operator in 4
dimensions, see (\ref{holo}) below with $m=0$, which immediately
generalizes to bifields. We leave this here as a conjecture, as
another remarkable feature related to the distinguished twist $D-2$
fields and their conservation laws. 

\subsection{An exotic restriction?} \label{exo}

The action of the group $SO(2,D)$ on the null cone $\xi\cdot\xi =
(\xi^0)^2 - (\xi^1)^2 - \dots - (\xi^D)^2 + (\xi^{D+1})^2=0$ in $D+2$
dimensions induces an action of $SO(2,D)/\ZZ_2$ on the projective cone
obtained by the identification $\xi\sim\lambda\xi$
($\lambda\in\RR\setminus\{0\}$). The projective cone is known as the
Dirac space or conformally compactified Minkowski spacetime $\overline
M_D\sim (S^1\times S^{D-1})/\ZZ_2$, into which $D$-dimensional
Minkowski spacetime is embedded as the chart  
$$x^\mu=\frac{\xi^\mu}{\xi^D+\xi^{D+1}} \qquad (\mu=0,\dots D-1),$$
so that $SO(2,D)/\ZZ_2$ becomes the conformal group. Restricting a 
4D conformal QFT to 2D, the relevant conformal group is $SO(2,2)/\ZZ_2\subset
SO(2,4)/\ZZ_2$, embedded as the subgroup that fixes the restricted
directions $2$ and $3$. This 2D conformal group is a direct product of
two M\"obius groups $SO(1,2) = SL(2,\RR)/\ZZ_2 =SU(1,1)/\ZZ_2$ acting on the
chiral variables $x^0\pm x^1$.  

There is another embedding of two commuting M\"obius subgroups
$SO(1,2)$ into $SO(2,4)/\ZZ_2$ as the subgroups that fix the
directions $0,1,2$ and $3,4,5$ respectively. One might wonder whether
this subgroup $G$ corresponds to some ``exotic'' 2D restriction. 

The first objection is that $G$ has no two-dimensional orbits in the
4D Dirac space $\overline M_4$, that could serve as the restricted 2D 
world hypersurface. But one could envisage a more abstract situation
following an idea of \cite{BS}: Let $\alpha^{(2)}_g$ denote the action
of $G$ on the 2D Dirac space $\overline M_2$, and fix any double cone
$O\subset \overline M_2$. Suppose we find a subalgebra $A$ on the
Hilbert space of the unrestricted 4D theory (where $G$ is unitarily
represented) with the properties that $U(g)AU(g)^*\subset A$ for all
$g\in G$ such that $\alpha^{(2)}_gO\subset O$, and $U(g')AU(g')^*$
commutes with $A$ for all $g'\in G$ such that $\alpha^{(2)}_{g'}O\subset O'$,
where $O'$ is the causal complement of $O$ in $\overline M_2$. In this
case, we may consistently {\em define}  
$$A(\alpha^{(2)}_gO):=U(g)AU(g)^*$$ 
for {\em all} $g\in G$. These algebras on the Hilbert space of the 4D
theory would then qualify as local algebras of a 2D CFT, satisfying
local commutativity, conformal covariance and isotony. The problem
with this is, however, that the $L^\pm_0$ generators of the embedded
subgroup do not have positive spectrum in the 4D representation --
which is related to the fact that their orbits in the 4D Dirac space
$\overline M_4$ are spacelike rather than future timelike. We shall
briefly return to this in Sect.\ \ref{sec5}.

\section{Conformal holography}\label{sec4}
\setcounter{equation}{0}

\subsection{Timelike surfaces}

The question arises to which extent one can recover a
$D$-dimensional QFT from its restrictions. Clearly, in some form the
higher-dimensional conformal symmetry group and its unitary
representation must be present in the lower-dimensional theory. It is
possible \cite{BN} to give a system of axioms on the inner symmetries
of a lower-dimensional GCI CFT, which ensure that the theory can be
extended to a higher-dimensional GCI CFT. 

\subsection{Lightfront holography} \label{lfhol}

The characteristic initial value problem for the Klein-Gordon operator in
$D>2$ dimensions consists in finding a solution to
$(\square_x+m^2)\varphi_m(x)=0$ with prescribed values
$\varphi_N(x_+,x_\perp)$ of $\varphi$ (as in Sect.\ \ref{null}) on the
null (characteristic) hypersurface $N=\{x\in\RR^4: x^0=x^1\}$ with
sufficiently rapid decay. 

A (unique?) solution is given in terms of the massive commutator function  
$$C_m(x-y)=  \int \frac{d^4k}{(2\pi)^{3}}
\;\delta(k^2-m^2)\mathrm{sign}(k^0) e^{-ikx}$$ 
by
\be \label{holo} 
\varphi_m(x) = -2i \int_N dy_+\,d^2y_\perp\;
C_m(x-y)\vert_{y_-=0}\;\varphi_N(y_+,y_\perp). \ee
Notice the fact that the kernel $C_m(x-y)\vert_{y_-=0}$ solves the KG
equation w.r.t.\ $x$, and restricts at $x_-=0$ to 
$$C_m(z)\vert_{z_-=0} = \frac 
i4\;\mathrm{sign}(z_+)\delta(z_\perp)\quad\RA\quad\partial_+
C_m(z)\vert_{z_-=0} = \frac i2\;\delta(z_+)\delta(z_\perp).$$
(\ref{holo}) not only solves the classical initial value
problem, but is indeed a relation between {\em quantum} fields in
different dimensions: namely, if one takes for
$\varphi_N(y_+,y_\perp)$ the chiral free field with two-point function
(\ref{lf2}) and computes the two-point function of the r.h.s.\ of
(\ref{holo}), one recovers the two-point function of the massive free
field in $\RR^4$.  

(\ref{holo}) is an adaptation of a similar formula used in \cite{DMP}
to pull back a state on the null future $\mathfrak{I}^+$ of an
asymptotically flat spacetime to a state on the bulk. The feature that
a (free) field in Minkowski spacetime can be reconstructed from its
restriction to the null hypersurface, which behaves like an
infinite-component chiral conformal field, was first pointed out by 
Schroer \cite{S1,S2}. 

Interestingly enough, the massive free field of {\em any mass} can be
recovered from the same conformal field theory on the lightfront,
given by the free currents $j_n(x_+)$ ($n\in\NN$), just by choosing
the mass in the commutator function $C_m$. Schroer calls this ``a
different 4D spacetime organization of the same quantum substrate''
(given by the chiral theory). Such a thing is possible because of the
universality of the separable ``inner'' Hilbert space $L^2(\RR^2)$.  

\subsection{2D boundary holography}

In two dimensions, the presence of a boundary at $x^1=0$ leads to a
reduction of the degrees of freedom because the boundary conditions
imply that the left- and right-moving chiral fields are no longer
independent but coincide with each other \cite{C,LR04}. In particular,
the restriction of the chiral fields to the time axis (= the boundary)
coincides with this chiral subtheory, while the restriction of
non-chiral fields (not satisfying 2D Huygens locality) will in general
be nonlocal on the time axis, but relatively local w.r.t.\ the chiral
subtheory. The full CFT in the Minkowski halfspace $x^1>0$ can be
recovered from the nonlocal boundary theory by a surprisingly simple
algebraic construction \cite{LR04}. 

Moreover, in a suitable state evaluated in the limit when all fields
are localized ``far away from the boundary'', the correlations
converge to those of an associated full 2D CFT with two independent
chiral subtheories \cite{LR09}. The basic mechanism that restores the
full 2D degrees of freedom (in particular, two chiral algebras) is the
decoupling of left and right movers in the limit under consideration,
due to the cluster property of the single chiral theory. The GNS
reconstruction from this factorizing state then produces the tensor
product of two chiral algebras.

\section{4D Positivity}\label{sec5}
\setcounter{equation}{0}
As mentioned before, the main open question concerning the double pole
solutions of Sect.\ \ref{bih} is, whether they are compatible with
Hilbert space positivity. To test positivity, one would like to split
correlation functions that should be positive by Hilbert space
positivity, into contributions that should be separately positive. 

Such a decomposition is the partial wave expansion: a given correlation
function splits into contributions 
\be\label{proj} 
\big\langle D(x_4) C(x_3)\Pi_\lambda B(x_2)A(x_1)\big\rangle 
\ee
where $\Pi_\lambda$ are projections onto the irreducible
representations $\lambda=(d,j_1,j_2)$ of the conformal group. Each
term (\ref{proj}) is a coefficient times a partial wave =
eigenfunction of differential operators corresponding to the three
Casimir operators (quadratic, cubic, and quartic). Positivity requires
in the simplest case, that all partial wave coefficients of
correlations of the type $\langle ABBA\rangle$ in (\ref{proj}) must be
nonnegative, and associated Cauchy-Schwarz (CS) inequalities
\cite{NRT05}. Even without knowing the six-point function, its mere
existence imposes via CS inequalities further nontrivial constraints
on the four- and two-point functions \cite{Y}. 

Let us notice here that positivity enters the analysis at several
stages. First of all, the fields of the theory are subject to the
unitarity bound \cite{M77}. Second, the condition that the operator
product expansion of two fields does not involve fields below the
unitarity bound, is reflected in bounds on the poles%
\footnote{ 
Concerning these bounds, there were some inaccuracies in the
admitted range of certain parameters around eq.\ (B.10) of
\cite{NRT05}. That the partial waves are regular and the expansion
formulae derived in \cite{NRT05} remain valid in the corrected parameter
range, was checked in \cite{IW}.} 
in the variables $\rho_{ij}$ \cite{NT01}, that were implicitly used
throughout Sect.\ \ref{sec2}. While we regard these bounds as
``kinematical'', the positivity of partial wave coefficients and the
associated CS inequalities are ``dynamical'' constraints which are
notoriously difficult to evaluate. 

It should therefore be clear that we can only test {\em necessary}
conditions for positivity througout. Even so, the partial wave
analysis is not practical for higher than four-point correlations,
because the computation of the 4D partial waves seems out of reach. We
therefore seek for simpler alternatives, that might give necessary
conditions for positivity.

\subsection{Positivity by restriction}
One option is to remark that restriction preserves Hilbert space
positivity, since it only amounts to limits in the test
function space, see also \cite{M}. Hence, a 4D double pole structure
must be rejected if its 2D restriction violates positivity.

Upon restriction, both the (tensor) fields will decompose into
(subtensor) fields, and the irreps will split into irreps of the
subgroup. Therefore, the restriction of a 4D partial wave is in 
general a sum of infinitely many 2D partial waves. To use this as a
tool, it is necessary to understand the branching rules.  

The branching of {\em representations} can be computed from
the characters $\chi(s,x,y) = \mathrm{Tr}\, s^{M_{05}}
(xy)^{M_{12}}(x/y)^{M_{34}}$ of the representations, counting the
multiplicities of the eigenvalues of the Cartan generators, see e.g.,
\cite{D}. For twist $\neq 2$   
$$\chi^{\mathrm{4D}}_{d,j_1,j_2}(s,x,y) =
\frac{s^{d}\cdot\chi_{j_1}(x)\chi_{j_2}(y)}{(1-sx^{\frac12}y^{\frac12})(1-sx^{\frac12}y^{-\frac12})(1-sx^{-\frac12}y^{\frac12})(1-sx^{-\frac12}y^{-\frac12})},$$
where $\chi_j(x)=x^{-j}+x^{1-j}+\cdots x^{j-1}+x^j$.
The restriction to 2D amounts to equating the parameters $x=y$. The
branching is then given by the expansion into 2D characters 
$$\chi^{\mathrm{2D}}_{h_+,h_-}(p,q)=\chi_{h_+}(p)\cdot\chi_{h_-}(q)=
\frac{p^{h_+}}{1-p}\cdot \frac{q^{h_-}}{1-q},$$
where $p=sx$ and $q=s/x$ couple to the chiral generators 
$L^{\pm}_0=\frac 12(M_{05}\pm M_{12})$. E.g., for the scalars
$j_1=j_2=0$, this gives the branching of representations
\be\label{repb}
D^{\mathrm{4D}}_{d,0,0}\big\vert_{\mathrm{2D}} = \bigoplus_{n} (n+1)\cdot
D^+_{(d+n)/2}\otimes D^-_{(d+n)/2}.\ee
Since the representations are generated by corresponding fields from
the vacuum, the multiplicity factor $n+1$ in this branching can be
easily understood as counting the derivatives of the field in the
restricted directions (leading to an increase of the dimension by one
unit), in accord with the rules obtained from \cite{BT,DMPPT}. In the
general case, the factor $\chi_j(x)^2$ produces more terms corresponding to
the decomposition of tensors into subtensors. 

At twist $d-j_1-j_2=2$ ($j_1,j_2\neq 0$), there are subtractions in
the characters reflecting the absence of some states due to the
conservation laws (\ref{cl}). The factor  
$\chi_{j_1}(x)\chi_{j_2}(y)$ has to be replaced by
$\chi_{j_1}(x)\chi_{j_2}(y) - s \chi_{j_1-\frac 12}(x)\chi_{j_2-\frac
  12}(y)$,
leading to a corresponding removal of some of the 2D subrepresentations.

The branching of {\em partial waves} follows a similar pattern. We
consider here only scalar fields. Since a restricted scalar field is
just another scalar field, only the decomposition of the projections
in (\ref{proj}) matters. For the most symmetric four-point case when
$A,B,C,D$ are scalars of the same scaling dimension $d$, we found the
following result.  

Extracting a prefactor $(\rho_{12}\rho_{34})^{-d}$, the 4D partial
waves depend only on the cross ratios
$s=\frac{\rho_{12}\rho_{34}}{\rho_{13}\rho_{24}}$,
$t=\frac{\rho_{14}\rho_{23}}{\rho_{13}\rho_{24}}$. For twist $2k$ 
and spin (tensor rank) $L=2j_1=2j_2$ of the
representation $\lambda$, they are given by
\cite{DO} 
$$\beta^{\mathrm{4D}}_{k,L}(u,v)=\frac{uv}{u-v}\big(G_{k+L}(u)G_{k-1}(v)-(u
\leftrightarrow v)\big),$$ 
where the ``chiral'' variables $u,v$ are
algebraic functions of $s,t$ given by $s=uv$, $t=(1-u)(1-v)$, and
$G_n(z)=z^n{}_2F_1(n,n;2n;z)$. Upon restriction to $D=2$, $u$ and $v$
become the chiral cross ratios $u=\frac{x_{12+}x_{34+}}{x_{13+}x_{24+}}$,
$v=\frac{x_{12-}x_{34-}}{x_{13-}x_{24-}}$. The 2D partial waves of
dimension $h_++h_-$ and helicity $h_+-h_-$ are
given by 
$$\beta^{\mathrm{2D}}_{h_+,h_-}(u,v) = G_{h_+}(u)G_{h_-}(v).$$
Using repeatedly the identity $\textstyle G_{n-1}(z)
-\frac{1-z/2}z\,G_n(z) + c_{n}\, G_{n+1}(z) =0$, 
where $c_{n}= \frac{n^2}{4(4n^2-1)}$, we found the recursion \cite{DM} 
$$\beta^{\mathrm{4D}}_{k,L} =
\sum_{m,n\geq0\atop m+n=L}\beta^{\mathrm{2D}}_{k+m,k+n} + c_{k+L}\,\beta^{\mathrm{4D}}_{k+1,L} + \sum_{\nu=1}^{[L/2]} (c_{k+L-\nu}-c_{k+\nu-1})\,
\beta^{\mathrm{4D}}_{k+\nu+1,L-2\nu}.
$$
For $L=0$ or $=1$, the last sum on the r.h.s.\ is empty. The 4D partial
waves on the r.h.s.\ can be iteratively expanded by the same formula,
giving all 2D partial waves of dimension $2k+L+2r$ in the $r$-th step
of the iteration: 
\bea\label{pwb} \beta^{\mathrm{4D}}_{k,0} = \sum_{r\geq 0}c_{k}c_{k+1}\dots
c_{k+r-1}\;\beta^{\mathrm{2D}}_{k+r,k+r},\\[-1mm]
\beta^{\mathrm{4D}}_{k,1} = \sum_{r\geq 0}c_{k+1}c_{k+2}\dots
c_{k+r}\Big(\beta^{\mathrm{2D}}_{k+r+1,k+r}+ \beta^{\mathrm{2D}}_{k+r,k+r+1}\Big).\nonumber\eea

If $L\geq 2$, the last sum contains negative coefficients (because
$c_n$ is mono\-tonously decreasing); but the iteration of the term
$c_{k+L}\,\beta^{\mathrm{4D}}_{k+1,L}$ contributes to the same 2D partial
waves, making the total coefficients positive, e.g.,
\bea \beta^{\mathrm{4D}}_{k,2} = \sum_{r\geq 0}c_{k+2}c_{k+3}\dots
c_{k+r+1}\cdot \Big(\beta^{\mathrm{2D}}_{k+r+2,k+r}+
\frac{c_{k+r+1}+c_{k+r}-c_k}{c_{k+r+1}}\,\beta^{\mathrm{2D}}_{k+r+1,k+r+1}
+\beta^{\mathrm{2D}}_{k+r,k+r+2}\Big).\nonumber \eea

Comparing (\ref{repb}) with (\ref{pwb}), there seems to be a
discrepancy, since the latter sum runs only over integer $r$, i.e.,
half of the representations present in (\ref{pwb}) are absent in the
restricted partial wave. This teaches us that in order to ``exhaust''
the full content of representations in a restricted partial wave, one
must also consider derivatives of the fields in the restricted
directions, before restricting.

In order to extend this tool to six-point functions, one would need to
know six-point partial waves. We do not know these partial waves, but
it is clear that the Casimir eigenvalue equations are much easier to
access in 2D than in 4D \cite{DM}. 

\subsection{The exotic restriction (continued)}

Let us resume the discussion of Sect.\ \ref{exo}. The generators
$L^\pm_0$ of the 2D conformal group embedded into the 4D conformal
group are, in this case, $M_{12}$ and $M_{34}$. Thus one should
obtain the decomposition of representations by putting $s=1$, and
letting $xy$ and $x/y$ play the role of $p$ and $q$ before. It is
then obvious that the expansion involves negative powers of $p$ and
$q$, reflecting the obvious fact that $M_{12}$ and $M_{34}$ do
not have positive spectrum in 4D positive energy representations. The
expansion technique of the previous subsection fails in this situation.

More detailed analysis of the spectrum of the two chiral Casimir
operators \cite{DM} indicates that the decomposition goes into a 
{\em continuum} of representations of the M\"obius groups with
positive and negative unbounded spectrum of $L^\pm_0$.  

\subsection{Characterization of twist 2 contributions}

Another idea to isolate parts from correlation functions that must be
separately positive, is to use the twist. This is a convenient
``quantum number'', but not an eigenvalue of any polynomial function
of the Casimir operators. Yet, as the discussion of biharmonic fields
shows, the projection to the sum of all twist 2 representations 
$$\big\langle \cdots \Pi_{\mathrm{twist}\;2} A'(x)A(y) \big\rangle  = 
\sum_{\lambda:\mathrm{twist}(\lambda)=2}\big\langle \cdots \Pi_\lambda A'(x)A(y) \big\rangle $$
is, after multiplication with $((x-y)^2)^{d-1}$, characterized by
the very simple pair of differential equations (\ref{bih}). 
This suggests the following potential technique. We know that
\be\label{VCBA}
\big\langle V(x,y) \Pi_{\mathrm{twist}\;2} C(x_3) B(x_2) A(x_1)
\big\rangle = \big\langle V(x,y) C(x_3) B(x_2) A(x_1) \big\rangle\ee
is a biharmonic function due to (\ref{bih}) for every biharmonic field
$V$. Since by conformal invariance, correlation functions depend
essentially only on the cross ratios, here regarded as ``collective
variables'', one may expect that the same information encoded in the
wave operators $\square_x$ and $\square_y$, can be encoded in a system
of differential operators w.r.t.\ the variables $x_1,x_2,x_3$,
annihilating $\langle VCBA\rangle$. Then, under the reasonable
hypothesis, that all biharmonic fields of the theory generate the
entire twist 2 subspace of the Hilbert space, this would imply that
the vector $\Pi_{\mathrm{twist}\,2}CBA\Omega$ solves the same
equations, and so does the six-point correlation function  
$$\big\langle A(x_6) B(x_5) C(x_4) \Pi_{\mathrm{twist}\;2} C(x_3) B(x_2) A(x_1)
\big\rangle.$$
This information can be used to compute the form of this contribution,
and to isolate the twist 2 part of a given six-point correlation
$\langle ABCCBA\rangle$, because the higher twists are less
singular. If the twist 2 part fails to be positive, the 
full six-point function is not positive. Ultimately, we would like to
apply this strategy to six-point double pole structures which appear in
correlations of the form $\langle VCCV \rangle$ \cite{NRT08}.

As a first step towards this program, we have tested the idea
on four-point functions \cite{IW}. So let $C$ be the unit operator in
(\ref{VCBA}). If $A$ and $B=A'$ have the same scaling dimension, it is
obvious that the twist 2 projection selects the biharmonic field
$V_{A'A}$, and it is also known that the wave operators w.r.t.\ $x$
and $y$, if expressed in terms of the cross ratios $s,t$, are the same
as the wave operators w.r.t.\ the arguments of $V_{A'A}(x_2,x_1)$. 
Hence, in this case the strategy works. 

Less obvious is the case when $d_A\neq d_B$.
The difference $d_B-d_A=2n$ must be even by GCI, and we may
assume $n>0$. Writing 
$$(x_{12}^2)^{d_A+n-1}\cdot\langle V(x,y) B(x_2) A(x_1)\rangle 
= f(x,y,x_2,x_1),$$  
we found \cite{IW} that biharmonicity in $x$ and $y$ implies the pair
of equations
$$\big[x_{12}^2\partial_1\!\cdot\!\partial_2 - 2(x_{12}\otimes
x_{12})\!\cdot\!(\partial_1\otimes\partial_2) + 2(n-1)x_{12}\!\cdot\!\partial_2 +
2(n+1)x_{12}\!\cdot\!\partial_1\big]f=0,$$
and 
$$\big(\partial_1^{\otimes n}\big)_{\rm traceless} f=0,$$
i.e., a pair of differential operators w.r.t.\ $x_1$ and $x_2$
characterizing ``twist 2'', as desired. The first equation is actually
equivalent to  
$$\langle V(x,y) (\mathcal{C}-\lambda)B(x_2)A(x_1)\rangle =0,$$ 
where $\mathcal{C}$ is the quadratic Casimir operator and $\lambda$
its eigenvalue in the scalar representation of dimension 2. Hence, the
twist 2 contribution $\Pi_{\mathrm{twist}\,2} B(x_2) A(x_1) \Omega$
consists of a scalar part only. This is an independent proof of Lemma
5.2 in \cite{NRT05} which states that the only twist 2 contribution in
the operator product expansion of two GCI scalar fields of different
dimension is the scalar $d=2$ representation. The second
equation is equivalent to the statement that every correlation
$(x_{12}^2)^{d_A+n-1}\langle \cdots \Pi_{\mathrm{twist}\;2} B(x_2) A(x_1)\rangle$ is a
homogenous {\em polynomial} in $\rho_{1i}$ of order $n-1$.  

An illustrating free field example for $n=2$ is the following. Let 
$\varphi$ be the massless free field, and $W_\mu$ a conformal vector
field of dimension $\Delta >3$. Then $A=\wick{W_\mu W^\mu}$ and
$B=\wick{[(\Delta-3)W^\mu\partial_\mu \varphi-\varphi(\partial_\mu W^\mu)]^2}$ are
conformal scalars of dimension $d_A=2\Delta$ and $d_B=2\Delta+4$. The
projection $\Pi_{\mathrm{twist}\,2}$ acting on $B(x_2) A(x_1) \Omega$
amounts to the contraction of all $W$ fields. The result is
$(x_{12}^2)^{-d_A-1}$ times the vector 
$$\big(x_{12}^2\square_2 - 4x_{12}\cdot\partial_2 + 8\big)\wick{\varphi^2(x_2)}\;\Omega$$
which is indeed annihilated by the two differential operators above. 
Splitting any correlation $(x_{12}^2)^{d_A+1}\langle \cdots
B(x_2) A(x_1) \rangle$ into a part in the kernel of the two differential
operators and a less singular part, uniquely selects this vector.
(Incidentally, in this case, the first operator is sufficient
to do the job.)

\section{Conclusion}

We have presented a number of ideas and new techniques which might be
developped into useful tools for the analysis of globally conformal
invariant correlation functions, especially the problem of Hilbert
space positivity of correlation functions that cannot arise from free
fields. Various side aspects, concerning the relations
between conformal QFT in four, two and one (chiral) dimensions were
also discussed.  

\bigskip

\noindent
{\bf Acknowledgements:} KHR thanks the Bulgarian Academy of Sciences
and the organizers of the Symposium ``Algebraic Methods in Quantum
Field Theory'', Sofia, May 15-16, 2009, for the invitation to this
event. He also thanks V. Moretti and B. Schroer for helpful comments
about aspects of lightfront holography.

\noindent
{\bf Note:} During the symposium, KHR became aware of the recent work
by G. Mack \cite{M}, also presented on that occasion, which has some
overlap with ours.

\end{document}